\documentclass[11pt,a4paper]{article}
\pdfoutput=1
\usepackage{jcappub}

\usepackage{xcolor}
\usepackage{bm}
\usepackage{subfigure}
\usepackage{feynmp}
\usepackage{extarrows}
\usepackage{slashed}
\usepackage{dcolumn}
\usepackage{verbatim}
\usepackage{here}
\usepackage{multirow} 
\allowdisplaybreaks[4]

\numberwithin{equation}{section}

\newcommand{\FR}[2]{\displaystyle\frac{\,{#1}\,}{#2}}
\newcommand{\fr}[2]{\mbox{$\frac{\,{#1}\,}{#2}$}}
\newcommand{\n}{\nonumber}

\def\bge{\begin{equation}}
\def\ede{\end{equation}}
\def\bga{\begin{aligned}}
\def\eda{\end{aligned}}
\newcommand{\beq}{\begin{equation}}
\newcommand{\eeq}{\end{equation}}
\newcommand{\bq}{\begin{equation}}
\newcommand{\eq}{\end{equation}}
\newcommand{\ba}{\begin{array}}
\newcommand{\ea}{\end{array}}
\newcommand{\beqa}{\begin{eqnarray}}
\newcommand{\eeqa}{\end{eqnarray}}
\newcommand{\beqs}{\begin{subequations}}
\newcommand{\eeqs}{\end{subequations}}

\def\({\left(}
\def\){\right)}

\def\End{\end{document}}
\newcommand{\order}[1]{\mathcal{O}({#1})}

\def\geqq{\geqslant}

\def\be{\beta}

\def\ep{\epsilon}
\def\lam{\lambda}
\def\rh{\rho}
\def\si{\sigma}

\def\bge{\begin{equation}}
\def\ede{\end{equation}}
\def\bga{\begin{aligned}}
\def\eda{\end{aligned}}
\def\bgp{\begin{pmatrix}}
\def\edp{\end{pmatrix}}
\def\bgs{\begin{subequations}}
\def\eds{\end{subequations}}
\def\di{{\mathrm{d}}}

\def\mh{m_h^{}}
\def\mt{m_t^{}}

\def\pd{\partial}

\def\wt{\widetilde}

\def\be{\beta}

\def\ep{\epsilon}

\def\lam{\lambda}
\def\rh{\rho}
\def\si{\sigma}

\def\Hu{{\mathcal{H}}}
\def\ns{n_s^{}}

\def\MP{M_{\text{P}^{}}}

\def\mutr{\mu_\text{tr}^{}}

\def\End{\end{document}}

\title{\huge Asymptotically Safe Higgs Inflation}

\author[a,b]{\large Zhong-Zhi Xianyu,}
\author[a,c,d]{\large~Hong-Jian He\,}

\affiliation[a]{Institute of Modern Physics and Center for High Energy Physics, \\
                Tsinghua University, Beijing 100084, China}
\affiliation[b]{Theoretical Particle Physics and Cosmology Group, Department of Physics, \\
                King's College London, London WC2R 2LS, UK}
\affiliation[c]{Center for High Energy Physics, Peking University, Beijing 100871, China}
\affiliation[d]{Kavli Institute for Theoretical Physics China, CAS, Beijing 100190, China}

\emailAdd{xianyuzhongzhi@gmail.com, hjhe@tsinghua.edu.cn}

\abstract{
\\[1mm]
We construct a new inflation model in which the standard model
Higgs boson couples minimally to gravity and acts as the inflaton.
Our construction of Higgs inflation incorporates the standard model with Einstein gravity
which exhibits asymptotic safety in the ultraviolet region.
The slow roll condition is satisfied at large field value due to the asymptotically safe
behavior of Higgs self-coupling at high energies.
We find that this minimal construction is highly predictive, and
is consistent with both cosmological observations and collider experiments.
}

\keywords{
\\[1mm]
Inflation, Quantum Gravity Phenomenology, Particle Physics\,-\,Cosmology Connection, \\
Cosmology of Theories Beyond the SM
\\[5mm]
JCAP (2014), final version [arXiv:1407.6993 [astro-ph.CO]]
}

\begin{document}

\maketitle

\setlength{\baselineskip}{18pt}

\setcounter{page}{2}
\vspace*{10mm}
\section{Introduction}
\label{sec:1}
\vspace*{2mm}

The cosmological inflation \cite{inf1}-\cite{inf5}
is a leading paradigm describing our universe in the pre-Big-Bang era.
Apart from solving a number of conceptual problems of Big-Bang cosmology,
it makes nontrivial predictions on the spectrum of primordial perturbation,
which can be directly compared with observations of cosmological microwave background
and large-scale structures. The inflation is expected to naturally happen
at ultra high energies, presumably around the grand unification (GUT) scale.

The released data of Planck satellite\,\cite{planck} last year
and the recent measurement of BICEP2 telescope\,\cite{BICEP2}
have generated excitements to test theories of inflation. More data are upcoming.
Indeed, these observations provide important information on the magnitude and
shape of the primordial perturbations through the CMB measurements with impressive accuracy.
By measuring the spectral index $\,n_s^{}\,$ of curvature perturbation, Planck has ruled out
the exactly scale invariant spectrum above $\,5\sigma\,$ level.
For the tensor-to-scalar ratio $\,r\,$,\, the Planck result is consistent with $\,r=0\,$,\,
while the BICEP2 result points to a larger value of $\,r\,$.\,
There is currently some tension between the indirect measurement of Planck on $\,r\,$
and the BICEP2 result, unless one assumes a rather large running of the spectral index  $\,n_s^{}\,$
(which would probably ruin the slow-roll approximation).
Meanwhile, the BICEP2 analysis is under further scrutiny, concerning its foreground contamination and
dust subtraction \cite{dust1,dust2,dust3,Planck-new}.
When adopting different dust models, the current BICEP2 result could be interpreted
as either a large $\,r\,$ or a vanishingly small $\,r\,$ plus foreground dust
contributions \cite{dust1,dust2,dust3}.
Both possibilities are consistent with the latest Planck measurement
on polarized dust emission \cite{Planck-new}.
Hence, at this stage, it is prudent to keep open-minded
on all possible values of $\,r\,$ for theory studies.

So far the Higgs inflation\,\cite{HI,HI2} appears
one of the most economical and predictive candidates
among all the proposed theories of inflation. It makes use of the discovered unique scalar particle ---
the Higgs boson \cite{Higgs2012,ICHEP2014}, as the inflaton, and does not assume any new particle beyond
the standard model (SM). Its key ingredient is to include the dimension-4 nonminimal coupling term
$\,\xi \mathcal{R}H^\dag H\,$  between the Ricci scalar $\,\mathcal{R}\,$ and Higgs doublet $\,H\,$
when combining the SM with Einstein gravity.
By taking account of radiative corrections, the Higgs inflation relates inflation parameters to
the masses of Higgs boson and top quark, which can be compared with both cosmological observations
and collider experiments \cite{HI_TwoLoop}. In particular, the requirement that the Higgs potential
to be positive up to the inflation scale puts a lower bound on Higgs mass $\,m_h^{}\,$  for a given value
of top mass $\,m_t^{}\,$,\, or equivalently, an upper bound on $\,m_t^{}\,$ for a given $\,m_h^{}\,$.\,
For instance, if we choose $\,\mh =126$\,GeV, then a positive Higgs potential requires $\,\mt\lesssim 171$\,GeV.
On the other hand, the latest measurements of Higgs mass
$\,\mh =125.6\pm 0.2(\text{stat})\pm 0.3(\text{syst})$\,GeV \cite{Higgs2014} and top mass
$\,m_t^{} = 173.39^{+1.12}_{-0.98}\,$GeV \cite{mt-new},\footnote{At the present, the most accurate
determination of top-quark mass comes from the world combination of
the ATLAS, CMS, CDF and D0 experiments\,\cite{Top2014},
$\,m_t^{} = 173.34 \pm 0.27(\text{stat}) \pm 0.71(\text{syst})\,$GeV.
This is given by the best fit to $\,m_t^{}\,$ as implemented in the respective Monte Carlo (MC) code,
and is normally called MC top mass. This MC mass definition can be converted to a theoretically well-defined
short-distance mass definition with an uncertainty of $\,\sim\!1\,\text{GeV}$\, \cite{mt-new},
and the resulted pole mass is derived\,\cite{mt-new},
$\,m_t^{} = 173.39^{+1.12}_{-0.98}\,$GeV.
The Snowmass study\,\cite{snowmass2013-top} showed that
the upgraded high luminosity LHC can further reduce the error $\,\Delta m_t^{}\,$ down to $500$\,MeV,
and an $e^+e^-$ lepton collider (like the ILC) would measure $\,\Delta m_t^{}\,$ to $100$\,MeV level.}.\,
together with the recent improved renormalization group analyses,
indicate that the stability of the SM Higgs potential is excluded above
$\,2\si\,$ level \cite{stability}. The Higgs self-coupling $\,\lam\,$ turns negative around
$\,10^{11}\,$GeV \cite{stability}, far below the expected inflation scale. This requires that
new physics should enter somewhere below $\,10^{11}$\,GeV.
It applies to all inflation models with inflation scale $\,V^{1/4}\,$ larger than $\,10^{12}$\,GeV,
including the Higgs inflation.
There exist different ways of extending the SM, with new physics ranging from the TeV scale up to
around $10^{11}$\,GeV. For instance, we can make a simple extension of Higgs inflation
with TeV scale new physics by adding only two weak-singlets (a real scalar and a vector-quark)
with masses of $\order{\text{TeV}}$ \cite{He:2014ora}.
Other extensions with new heavy particles at various energy scales
were also studied recently \cite{ExtHI2}.

In this work, we explore a novel possibility that {\it the new physics is provided by quantum gravity itself,}
without introducing any new particle beyond the SM. Although a complete quantum theory of gravity remains
uncertain due to the perturbative nonrenormalizability of Einstein general relativity, there are strong
indications that the general relativity may flow nonperturbatively
to a nontrivial ultraviolet (UV) fixed point,
a scenario called the asymptotic safety (AS) as proposed by Weinberg \cite{AS}, and
further developed by many others \cite{ASReview}.
In this scenario, all dimensionless couplings flow to constants in the UV, leading to
a nontrivial conformal field theory (CFT).
For our purpose, we will study a new Higgs inflation with AS as the UV completion of the model.
In the literature, many applications of the AS to the SM and cosmology
were considered \cite{AS_HiggsMass}-\cite{ASInf-3}.
These include such as the constraints on the SM Higgs mass under AS \cite{AS_HiggsMass,Sannino},
the pure AS inflation without inflaton \cite{ASInf-1,ASInf-2},
and certain conventional inflation models improved by AS \cite{ASInf-3}, etc.

For our study, we will construct a new inflation model implementing the AS scenario,
with the SM Higgs boson acting as the inflaton,
which may be called asymptotically safe Higgs inflation (ASHI).
A crucial difference of the ASHI from the conventional Higgs inflation is that {\it it does not require
the nonminimal coupling $\,\xi R H^\dag H$,\, and the inflation is driven by an AS improved Higgs potential.}
Since no large nonminimal coupling is present, the model is free from the potential unitarity constraints
in the conventional Higgs inflation \cite{HI_naturalness}.
An ingredient of our ASHI is the conjecture suggested by previous studies\,\cite{AS_HiggsMass}
that the Higgs self-coupling $\,\lam\,$ and its beta function $\,\beta_\lam^{}\,$
approach zero at a transition scale $\,\mutr\,$,\,
around and above which the AS effects become important.
Then, the resultant Higgs potential is uniquely determined by inputting the Higgs and top masses
$\,(\mh,\,\mt)\,$  measured at weak scale, which is highly predictive.
As we will show, the Higgs potential that reproduces the observed curvature perturbation requires
$\,(\mh,\,\mt)\,$  to lie within a narrow strip, perfectly consistent with
the current collider measurements\,\cite{Higgs2014,mt-new}.

\vspace*{2mm}
\section{Asymptotic Safety and Running Couplings}
\label{sec:2}
\vspace*{2mm}

To motivate a Higgs inflation via the AS approach (without nonminimal coupling),
we start with a brief review on the relevant ingredients of the AS.
More details are given in the nice reviews \cite{ASReview}\,\footnote{The
existence of UV fixed points for AS was first proposed by S.\ Weinberg as a conjecture
\cite{AS}.  Since then, there have been extensive studies on it
and accumulating evidences were found in support of the AS \cite{ASReview}.
With truncated effective action, it has been shown that the UV fixed point of Einstein-Hilbert action
remains intact after adding higher order operators. So the AS with UV fixed point, though not yet
a rigorously proven fact by mathematics (due to the high nonlinearity of functional RG equations),
provides a serious possibility for the UV completion of general relativity.}.
The starting point of AS is a truncated effective action $\,\Gamma(\mu)\,$
as a direct generalization of the Einstein general relativity,
 \beqa
 \Gamma(\mu) \,=\,
 \FR{M_\text{P}^2(\mu)}{2}\!\int\!\!\di^4x\,\sqrt{-g}\,\big[\mathcal{R}-2\Lambda(\mu)\big] +\cdots,
 \eeqa
 where $M_\text{P}(\mu)$ and $\Lambda(\mu)$ are the running Planck mass and cosmological constant,
 and the dots ``$\cdots$'' represent gauge-fixing term, ghost term, and possible terms of matter fields.
 The scale dependence of running couplings are governed by the beta functions through
 $\,\di \MP(\mu)/\di\ln\mu = \be (\MP,\Lambda,\cdots)$, etc., where the dots denote matter couplings.
 Here the $\beta$ functions can be derived by solving the functional renormalization group equation
 of the effective action $\,\Gamma(\mu)$,\, and detailed calculations have revealed the existence of
 UV fixed points for dimensionless Planck parameter $\,\wt{M}_\text{P}^2(\mu)\equiv \mu^{-2}M_\text{P}^2(\mu)$,\,
 dimensionless cosmological constant $\,\wt\Lambda(\mu)\equiv \mu^{-2}\Lambda(\mu)$,\,
 and dimensionless matter couplings (when present).
 The fixed-point values $\wt M_\text{P*}$ and $\wt \Lambda_*$ are nonzero,
 implying a nontrivial interacting CFT in the UV limit.
 Hence, there exists a transition scale $\,\mutr\,$ from the low energy general relativity
 to a conformal phase \cite{ASReview,AS_HiggsMass}.
 Below $\,\mutr\,$ the dimensionful Planck mass $\,\MP\,$
 and cosmological constant $\,\Lambda\,$ remain roughly scale-independent,
 while above $\,\mutr\,$,\,  one gets the power-law running,
 $\,\MP (\mu)\sim \mu \wt{M}_\text{P*}^{}\,$ and
 $\,\Lambda(\mu)\sim \mu^2 \wt{\Lambda}_*^{}$\,.\,
 The transition scale for pure gravity sector is somewhat lower than the conventional Planck scale
 $\,M_{\text{P0}}^{} \equiv M_{\text{P}}^{}(\mu \!=\! 0) =(8\pi G)^{-1/2}\simeq 2.4\!\times\! 10^{18}$\,GeV,\,
 and will become model-dependent if matter fields are included.

 With the nonperturbative AS scenario,
 one also obtains similar UV behaviors for matter couplings \cite{ASReview,AS_HiggsMass,AFg0,AFg}.
 The gravitational contributions to the beta function for a given matter coupling $\,\lam_j^{}\,$
 typically takes the form \cite{AS,AS_HiggsMass},
 \beqa
 \label{beta_grav}
 \beta_j^{\text{grav}}(\lam_j^{}) \,=\,
 \FR{a_j^{}}{\,8\pi\,}\FR{\mu^2}{\,M_\text{P}^2(\mu)\,}\lam_j^{}\,,
 \eeqa
 where $\,\lam_j^{}\,$ can be Higgs self-coupling, gauge couplings and Yukawa couplings in the SM.
 These dimensionless matter couplings exhibit power-law running in general \cite{ASReview,AS_HiggsMass}.
 In the literature\,\cite{ASReview,AS_HiggsMass,AFg0,AFg}
 there are various detailed analyses for deriving the coefficients $\,a_j^{}\,$
 via nonperturbative approaches, which indicate the existence of Gaussian UV fixed points
 ($\lam_{j*}^{}\!=0$) for the gauge couplings and matter couplings under the AS
 (though other possible nontrivial UV fixed points may exist as well).
 Motivated by the AS scenario and following the conjecture of Ref.\,\cite{AS_HiggsMass}, we will choose
 Gaussian UV fixed points ($\lam_{j*}^{}\!=0$) with $\,a_j< 0\,$ for our current setup.
 On the other hand, it is known that for the SM with 
 $\,(\mh ,\,\mt)\simeq (126,\,173)$\,GeV,
 the Higgs self-coupling $\,\lam\,$ reaches zero at a scale $\,\mu_0^{}\ll \MP$,\,
 and becomes negative above $\,\mu_0^{}$\,.\, This is due to the negative contribution of
 the top-Yukawa coupling to the SM $\beta$-function,
 which signals instability of the SM Higgs potential \cite{stability}.

 \begin{figure}
   \centering
   \includegraphics[width=10.5cm,height=8.5cm]{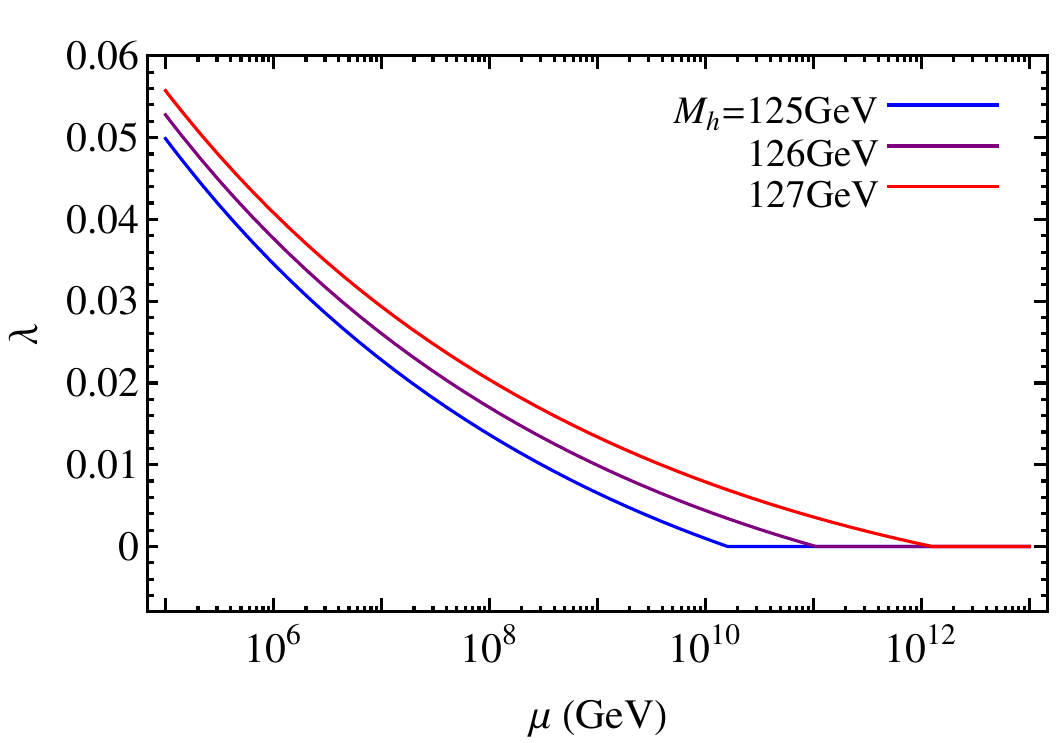} 
   \caption{Running Higgs self-coupling $\,\lam\,$ as a function of the energy scale $\,\mu\,$,\,
   with the asymptotic safety dynamics setting in at scales $\,\mu\geqq\mu_0^{}\,$.\,
   The (blue,\,purple,\,red) curves from bottom to top correspond to the Higgs mass
   $\,\mh =(125,\,126,\,127)$\,GeV, respectively.
   The top quark mass is taken to be $\,\mt =173.3$\,GeV.}
   \label{fig:1}
   \vspace*{5mm}
 \end{figure}

 When gravitational and SM $\beta$-functions are considered together,
 it was noted \cite{AS_HiggsMass} that the Higgs self-coupling $\,\lam\,$
 and its beta function $\,\be_\lam^{}$\, could remain vanishing once $\,\lam\,$ reaches zero,
 due to a Gaussian UV fixed point induced by the nonperturbative quantum-gravity corrections of AS.
 In Ref.\,\cite{AS_HiggsMass}, such considerations were put in use to predict
 the SM Higgs mass in the context of AS scenario,
 where the AS occurs at the conventional Planck scale and inflation was not considered.
 Now, we can identify the recently discovered 126\,GeV particle\,\cite{Higgs2012,ICHEP2014}
 as the SM Higgs boson.
 Thus, the scale $\,\mu_0^{}\,$ at which $\,\lam\,$ vanishes is around $\,10^{11}$\,GeV
 as derived from the SM two-loop $\beta$-functions.
 So we are led to {\it the conjecture that the transition scale
 $\,\mutr\,$ of AS may be around} $\,\order{\mu_0}\sim 10^{11}$\,GeV\,
 due to the nonperturbative dynamics of quantum gravity,
 and both $\,\lam (\mu)\,$ and $\,\be_\lam^{}\,$ will remain zero
 at all scales above $\,\mu_0^{}\,$ due to the Gaussian UV fixed point of AS.
 It is quite possible that the scale of quantum gravity may appear much below the Planck scale,
 as it does in many well-motivated extra dimensional models.
 For physical applications, the collider phenomenology of TeV scale AS scenario was studied
 before\,\cite{ASappx} and recently in the context of spontaneous dimensional reduction \cite{SDR}.
 Under the conjecture that  $\,\lam\,$ and $\,\be_\lam^{}\,$ approach zero at and above $\,\mu_0^{}$\,
 due to the UV fixed point induced by the AS dynamics,
 it follows directly that the transition scale $\,\mutr\,$ and Higgs mass
 $\,\mh\,$ are correlated through the requirement of $\,\mutr = \order{\mu_0^{}}$.\,
 In fact, for scales $\,\mu < \mu_0^{}\sim 10^{11}$\,GeV,
 the nonperturbative quantum gravity contribution (\ref{beta_grav}) to $\,\beta_\lam^{}\,$
 can be safely neglected, and thus $\,\mu_0^{}\,$ is fully determined
 by $\,(\mh,\,\mt)\,$ and the SM $\beta$-functions.
 (Hence, for our practical analysis, except the existence of the UV fixed point
 at the scale $\,\mutr$ via gravity-induced $\beta$-function (\ref{beta_grav}),
 our study does not rely on any detail of solving functional RG equations of AS
 at scales above $\,\mutr$\,.)

\vspace*{2mm}

 In Fig.\,\ref{fig:1}, we present the two-loop running of Higgs self-coupling $\,\lam(\mu)\,$
 to illustrate the transition from the conventional SM running behavior at low energies
 up to the phase of vanishing $\,\lam\,$.\, The renormalization group equations for such running
 coincide with that of the SM for the scales $\,\mu < \mutr\,$,\,
 which we summarize in Appendix\,\ref{app}.
 As a first approximation for phenomenological studies, we model this transition to happen abruptly
 at the scale $\,\mu_0^{}\,$.\,  One may improve this approximation by considering a more elaborated model
 dealing with the detailed nonperturbative dynamics of AS,
 but for our present purpose of demonstration
 it is enough to take the abrupt transition.
 In Fig.\,\ref{fig:1}, the three curves from bottom to top correspond to the Higgs mass
 $\,\mh =(125,\,126,\,127)$\,GeV, respectively. We also input the top quark mass $\,\mt =173.3$\,GeV.\,
 For practical analysis, we will take $\,\mutr = C\mu_0^{}\,$,\,
 with the dimensionless ratio $\,C = \order{1}$.\,
 It is reasonable to conjecture\,\cite{AS_HiggsMass}\cite{AFg0} that all the matter couplings
 (Higgs self-coupling, gauge couplings, Yukawa couplings) become vanishingly small
 around the transition scale $\,\mutr\,$,\, according to the discussions above.
 We will not illustrate the quantitative running behavior of gauge and Yukawa couplings here,
 since these technical details are not needed for our following analysis of Higgs inflation.

\vspace*{2mm}
\section{Asymptotically Safe Higgs Inflation}
\label{sec:3}

 In this section, applying the asymptotical safety scenario discussed in Sec.\,\ref{sec:2},
 we construct a new Higgs inflation model which minimally couples Higgs boson to the Einstein gravity.
 For convenience, we may denote this as asymptotically safe Higgs inflation (ASHI).
 Our model requires no new particles beyond the SM, except a well-motivated conjecture that
 the Einstein gravity exhibits asymptotic safety\,\cite{AS,ASReview},
 under which the running Higgs self-coupling $\,\lam\,$  behaves as in Fig.\,\ref{fig:1}
 and other matter couplings approach zero around and above the transition scale
 $\,\mutr\sim\mu_0^{}$\, \cite{AS_HiggsMass,AFg0}.

 The standard Friedmann-Robertson-Walker (FRW) cosmology remains unchanged in the AS scenario
 except that the Planck mass becomes scale-dependent,
 \beqa
 \label{eq:MP-run}
   M_\text{P}^2(\mu) \,=\, M_\text{P0}^2\bigg(1+\FR{\mu^2}{\mu_\text{tr}^2}\bigg),
 \eeqa
 where  $\,M_{\text{P0}}^{}=(8\pi G)^{-1/2}\simeq 2.4\!\times\! 10^{18}$\,GeV
 is the conventional reduced Planck mass.
 When applying the formula \eqref{eq:MP-run} to cosmology,
 one needs to properly choose the scale $\mu$\,.\,
 A reasonable choice of this scale should be such that the corrections from both radiative correction
 and higher dimensional operators are controllably small.
 To find the optimal choice for $\,\mu$\,,\,
 we first note that the one-loop corrections to the Higgs potential take the form,
 \beqa
   \label{1loopV}
   \Delta V \,=\, \sum_j \FR{\,z_j^{} M_j^4\,}{16\pi^2}\log\FR{M_j^2}{\mu^2} \,,
 \eeqa
 where the sum is taken over all SM fields interacting with the Higgs field $\,h$\,,\,
 the parameters $\,\{z_j^{}\}\,$ are $\,\order{1}$\, coefficients, and
 $\,\{M_j^{}\}\,$ are effective masses of the fields running in the internal loop.
 The effective mass $\,M_j^{}\,$ depends on the background Higgs field $\,h\,$
 and background spacetime curvature $\,\mathcal{R}\sim \Hu^2\,$.\,
 So it takes the following form,
 \beqa
   \label{EffMass}
   M_j^2 = \order{\lam_j^{}}\,h^2 + \order{1}\,\Hu^2 \,,
 \eeqa
 where $\,\lam_j^{}\,$ is the coupling of the internal field with $\,h$,\,
 and $\,\Hu\,$ is the Hubble parameter.

 In the conventional Higgs inflation, matter couplings are roughly of $\order{1}$,\,
 and the background Higgs field $\,h\,$ is much larger than the Hubble parameter
 $\,\Hu\,$ during the inflation.
 Thus,  to minimize radiative corrections (\ref{1loopV}) for the Higgs potential,
 it is natural to choose $\,\mu =h\,$.\footnote{In the conventional Higgs inflation,
 due to the presence of nonminimal coupling term $\,\xi {\cal R}H^\dag H$\,,\,
 two different approaches exist\,\cite{HI_TwoLoop}:
 one is in Jordan frame, and the other is in Einstein frame
 (which has the nonminimal coupling term transformed away).
 The choice $\,\mu =h\,$ is taken for the Jordan frame analysis,
 while in Einstein frame one chooses $\,\mu=h/\Omega\,$ instead,
 where $\,\Omega\,$ is the Weyl factor that brings the spacetime metric
 from Jordan frame to Einstein frame.}\,
 But, in the present case of ASHI, all matter couplings $\,\lam_j^{}\,$
 become vanishingly small around the scale $\,\mu_0^{}\,$ (due to the Gaussian UV fixed point),
 as explained in Section\,\ref{sec:2}. Thus, on the right-hand-side of (\ref{EffMass}),
 the second term dominates over the first term during inflation.
 As a result, we should choose $\,\mu =\Hu\,$ instead.

 Besides, we further check the contributions from possible higher dimensional operators and make sure
 that they are also sufficiently suppressed, since these operators are expected to appear
 due to the effect of AS. The contributions to the effective action from such AS-induced
 higher dimensional operators could take the form,
 \beqa
   \Delta\Gamma ~=\,
   \sum_{n\geqq 2}\FR{w_n^{}}{~\mu_\text{tr}^{2(n-2)}~}\mathcal{R}^n \,,
 \eeqa
 where $\,\{w_n^{}\}\sim\order{1}$ are dimensionless coefficients,
 and for notational simplicity we use $\,\mathcal{R}^n\,$ to generically denote
 all possible contractions among $\,n\,$ factors of Riemann curvature tensor
 $\,\mathcal{R}_{\mu\nu\rh\si}^{}\,$.\,
 During the inflation, we have $\,\mathcal{R}\sim \Hu^2$\,
 and $\,\Hu < \mu_0^{}\,$,\, as will be shown below.
 Recalling that $\,\mutr = C\mu_0^{}\,$ with $\,C=\order{1}\,$,\,
 we can thus estimate the contributions of higher dimensional operators to be
 $\,\order{\Hu^{2n}/\mu_0^{2(n-2)}} \lesssim \order{\Hu^4}\,$.\,
 This should be compared with the leading term in the gravitational sector,
 namely the Einstein-Hilbert action,
 $\,M_\text{P}^2(\mathcal{H})\mathcal{R}\sim M_{\text{P0}}^2\Hu^2$.\,
 Hence, it is clear that these higher order contributions can be safely neglected
 due to  $\,\Hu^2 \ll M_{\text{P0}}^2$\,.
 We also note that there could be higher dimensional operators containing matter fields
 in addition to gravitational field.
 Contributions of such operators are suppressed
 by the approximate shift symmetry of the effective Lagrangian, as in all sensible
 large field inflation models. In our case, the shift symmetry may be a low energy
 consequence of local conformal symmetry at the UV fixed point.

 So far we have justified the natural choice of the renormalization scale $\,\mu=\Hu\,$.\,
 Then, we can perform the inflationary analysis for the ASHI model in a straightforward way.
 The standard slow roll paradigm will apply directly,
 where the first two slow-roll parameters are given as follows,
 \beqa
 \ep \,=\, \fr{1}{2}M_\text{P}^2(\Hu)(V'/V)^2, ~~~~~
 \eta \,=\, M_\text{P}^2(\Hu)V''/V \,,
 \eeqa
 where $\,V=\fr{1}{4}\lam(\Hu) h^4$\, is the Higgs potential in the unbroken phase of electroweak symmetry,
 and its derivatives are defined by
 $\,V'\equiv\pd V/\pd h\,$ and  $\,V''\equiv \pd^2 V/\pd h^2$\,.\,
 The slow-roll condition is as usual, namely,  $\,\ep <1\,$ and $\,|\eta|<1\,$.\,
 Even before doing a detailed calculation, we can immediately realize a plateau-like behavior of $\,V\,$
 at large $\,h\,$.\,  In fact, the sliding scale, namely the Hubble parameter $\,\Hu$\,,\,
 increases monotonically with $\,h\,$.\, On the other hand, the running coupling $\,\lam(\Hu)$\,
 decreases monotonically with $\,\Hu\,$ for $\,\Hu<\mu_0^{}\,$.\,
 Thus, we have two competing factors: the $\,h^4\,$ drives $\,V\,$ upwards,
 while $\,\lambda(\Hu)\,$ drives $\,V\,$ downwards.
 This competition must end in a tie, since $\,\Hu\,$ can never exceed $\,\mu_0^{}\,$
 [otherwise $\,\lam\,$ would vanish and drive $\,\Hu\,$
 to very small values, cf.\ Eq.\,\eqref{eq:Feq} below].
 The scalar potential $\,V\,$ should keep increasing with $\,h\,$,\,
 but be fairly flat when $\,h\,$ becomes large.

 \begin{figure}[t]
   \centering
   \includegraphics[width=11cm,height=8.5cm]{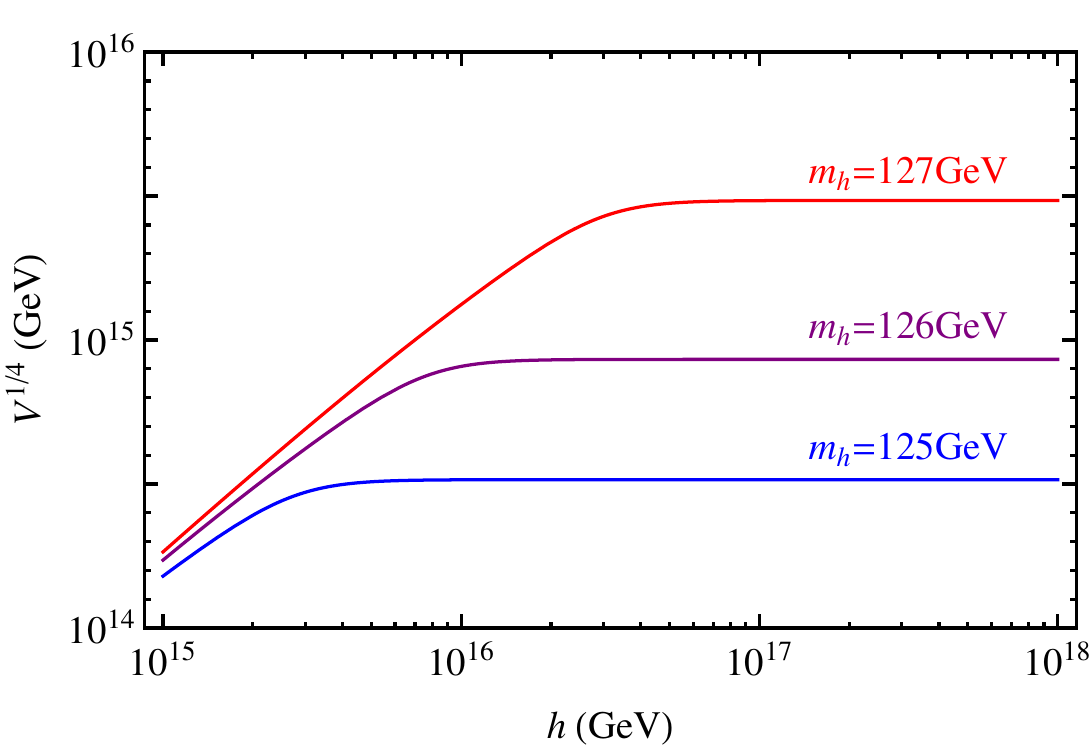}
   \vspace*{-1.5mm}
   \caption{Higgs potential $\,V=V(h)\,$ with slow-roll approximation at large $\,h\,$.\,
   The (blue,\,purple,\,red) curves from bottom to top correspond to Higgs mass
   $\,\mh =(125,\,126,\,127)$\,GeV, respectively. The top mass is taken to be $\,\mt=173.3$\,GeV.}
   \label{fig:2}
   \vspace*{7mm}
 \end{figure}
 \begin{figure}[h]
   \centering
   \includegraphics[width=11cm,height=8.5cm]{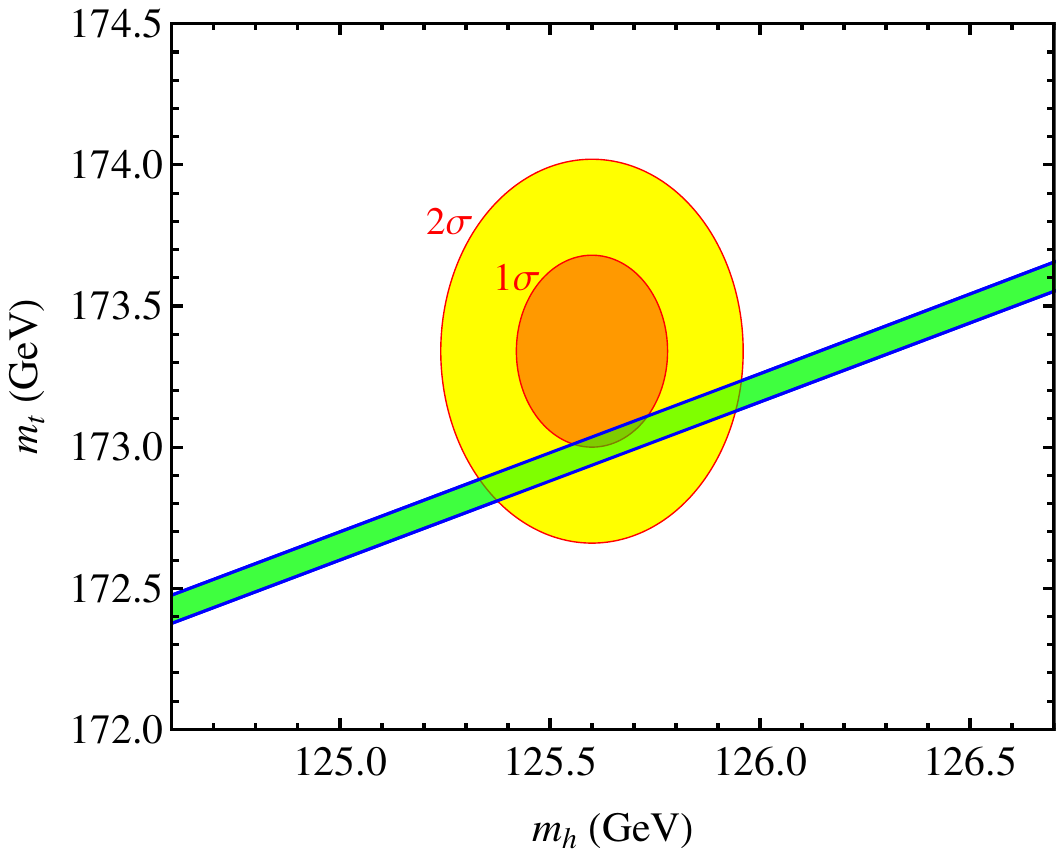} 
   \vspace*{-1.5mm}
   \caption{Predictions of top mass $\,\mt\,$ and Higgs mass $\,\mh\,$ by the
    asymptotically safe Higgs inflation (ASHI).
    The green strip shows the range consistent with cosmological data, while the orange and yellow
    regions depict the current constraints on $\,(\mh,\,\mt)\,$ by the collider experiments
    at $\,1\sigma\,$ and $\,2\sigma\,$  confidence levels, respectively.}
   \label{fig:3}
 \end{figure}

 The above analysis shows that there must be a region with large $\,h\,$
 where the slow-roll approximation holds.
 Let us focus on this case, and we have the simplified Friedmann equation,
 \beqa
 \label{eq:Feq}
   3M_\text{P}^2(\Hu)\Hu^2 \,=\, V \,=\, \fr{1}{4}\lam(\Hu)\,h^4 \,.
 \eeqa
 Below the transition scale $\,\mutr\,$,\, we can use the two-loop SM $\beta$-functions
 to derive the running behavior of $\,\lam(\Hu)\,$  as shown in Fig.\,\ref{fig:1}.
 Thus, we can solve the Hubble parameter $\,\Hu=\Hu(h)$\, from Eq.\,\eqref{eq:Feq}
 as a function of Higgs field $\,h\,$,\, and substitute it into the potential $\,V$,\,
 which then becomes $\,V=\fr{1}{4}\lam\big(\Hu(h)\big)h^4$\,,\,
 or equivalently, $\,V=3M_\text{P}^2\big(\Hu(h)\big)\Hu^2(h)$\,.\,
 With these, we can plot the inflation potential $\,V\,$ in Fig.\,\ref{fig:2}
 with different choices of Higgs mass $\,\mh\,$.\, As we see,
 Fig.\,\ref{fig:2} demonstrates that the potential is extremely flat
 when $\,h\gtrsim 10^{16}$\,GeV.

 With this inflation potential, we can perform standard calculation of slow-roll parameters.
 There is one free parameter in our model, namely, the ratio $\,C =\mutr /\mu_0^{}\,$,\,
 which is of $\,\order{1}$\,.\,  We find that the model fits the data well for $\,C\gtrsim 1\,$.\,
 The end of the inflation is indicated in our case by $\,|\eta|=1$\,.\,
 Then, taking the number of $e$-folding $\,N_e=50-60$\, for the inflation,
 we can derive the slow-roll parameters at the beginning of the observable inflation.
 These include the amplitude $\,V/\ep\,$,\,
 the scalar spectral index $\,n_s^{}= 1-6\ep +2\eta\,$,\,
 and the tensor-to-scalar ratio  $\,r = 16\,\ep\,$.

 \begin{figure}
   \centering
   \includegraphics[width=11.7cm,height=9.6cm]{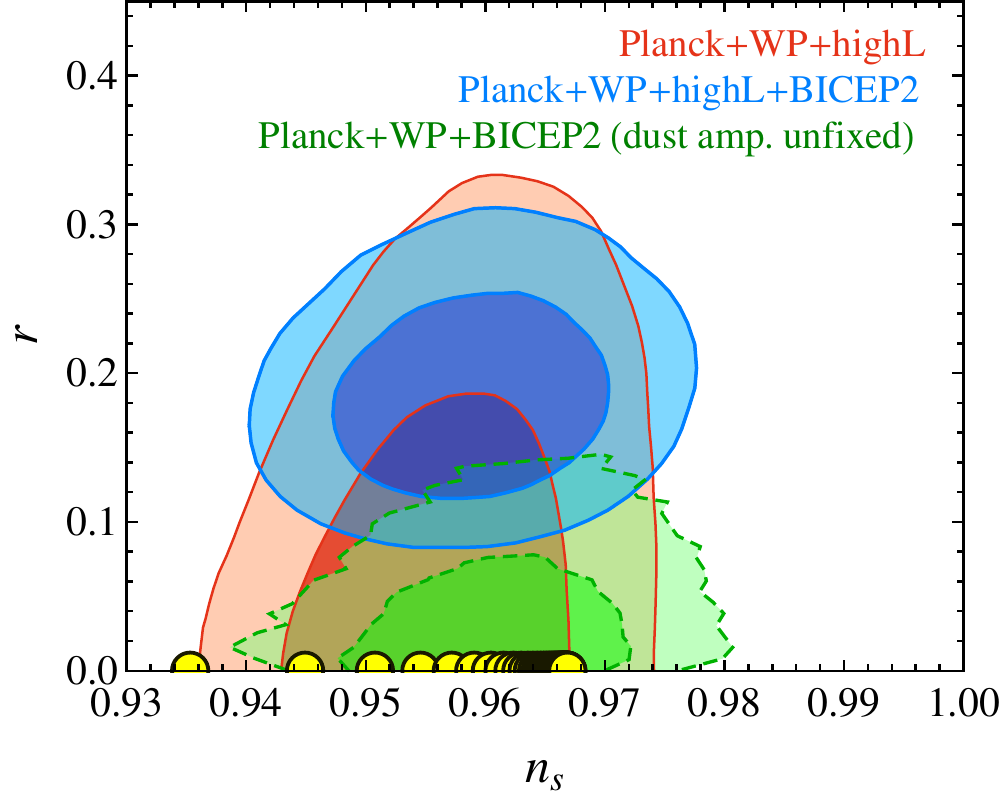}
   \vspace*{-1.5mm}
   \caption{Predictions of asymptotically safe Higgs inflation (ASHI) on the spectral index $\,\ns\,$
   and tensor-to-scalar ratio $\,r\,$.\, The yellow dots represent predictions of the ASHI with the ratio
   $\,C=\mutr/\mu_0^{}=1,\,1.2,\,1.4,\,\cdots,\,10$,\,  from left to right with step-width
   $\,\Delta C=0.2\,$.\, The red and blue shaded regions are observed limits at 68\%C.L. and 95\%C.L.,
   taken from Fig.\,13 of Ref.\,\cite{BICEP2}. The green region is taken from Fig.\,1 of Ref.\,\cite{dust2},
   in which the amplitude of dust polarization is not assumed \emph{a priori.} }
   \label{fig:4}
 \end{figure}

 In our model, the height of the inflation potential, and thus the amplitude of the curvature perturbation,
 is rather sensitive to initial values of $\,\mh\,$ and $\,\mt\,$,\,
 as can be seen from Fig.\,\ref{fig:2}.  Hence, this puts a nontrivial constraint on the range of
 \,$(\mh,\,\mt)$\, to fit the observed value of $\,(V/\ep )^{1/4}$\,
 by Planck satellite, which is
 $\,(V/\ep )^{1/4} \,=\,
 \( 0.0270^{+0.0010}_{-0.0009}\) \!\MP
 $
 at $2\sigma$ level \cite{PlanckCosPara}.
 In Fig.\,\ref{fig:3}, we present this constraint in the plane of \,$(\mh,\,\mt)$.\,
 The green strip represents the viable parameter region allowed by Planck data at $\,2\si\,$ level,
 and with the variation of $e$-foldings within $\,N_e=50-60$\,.\,
 In the same plot, we further display the current constraints on $\,(\mh,\,\mt)$\,
 by the collider experiments\,\cite{Higgs2014,Top2014},
 at $\,1\,\sigma\,$ (orange) and $\,2\,\sigma\,$ (yellow) confidence levels, respectively.
 In this plot, we take a sample input $\,C=2\,$ for demonstration.
 It is evident that the ASHI model fits well with both the Planck data and collider measurements.

  On the other hand, we find that the prediction of $\,(n_s^{},\,r)$\, is not sensitive to
  the initial values of Higgs and top masses at weak scale.
  Hence, we take the sample inputs $\,(\mh,\,\mt)=(125.6,\,173.0)$\,GeV, and set the number of $e$-folding
  $\,N_e=50$\,.\,  We present our predictions of \,$(\ns,\,r)$\, in Fig.\,\ref{fig:4}.
  In this plot, we vary the scale ratio $\,C=\mutr /\mu_0^{}\,$ over the range of
  $\,C=1-10\,$ with a step-width
  $\,\Delta C=0.2\,$.\,  We have derived the predictions for the spectral index $\,\ns\,$,\,
  which ranges from $\,\ns =0.935\,$ (for $\,C=1\,$) to $\,\ns =0.967\,$ (for $\,C=10$\,).\,
  We deduce the predicted tensor-to-scalar ratio $\,r = \order{10^{-7}}$\, in all cases.
  These predictions are depicted by the yellow dots in Fig.\,\ref{fig:4}.
  We see that the predicted value of $\,\ns\,$ increases quickly for $\,C\gtrsim 1\,$,\,
  and converges effectively around $\,r=0.967\,$ when $\,C>6\,$.

 For comparison, we further present constraints from cosmological observations in the same Fig.\,\ref{fig:4}.
 The red and blue contours are taken from Ref.\,\cite{BICEP2}, showing the measurements of $\,(\ns,\,r)$\,
 at $\,k=0.002\,\text{Mpc}^{-1}$.\, The green contours are taken from Ref.\,\cite{dust2},
 where no assumption on the amplitude of foreground dust polarization is made, except for its scaling behavior.
 When the dust amplitude is allowed to vary, the analysis of Ref.\,\cite{dust2} shows
 that the joint fit to Planck, WMAP and BICEP2 is consistent with $\,r=0\,$.\,
 It should be noted that the analysis of Ref.\,\cite{dust2} was done at $\,k=0.05\,\text{Mpc}^{-1}$.\,
 Hence, one should not compare this result (green contours) directly with that of Planck (red contours).
 Nevertheless, we present them in the same plot to show that our model with $\,C\gtrsim 1\,$
 can provide a successful inflation and fit well with the current measurements. The upcoming data
 from Planck, Keck Array and other $B$-mode measurements will further pin down the issue of potential
 foreground contamination, and thus provide more effective tests on the ASHI model.

 It is also useful to compare the inflation predictions of our ASHI model with
 that of related inflation models, including the conventional Higgs inflation
 with large non-minimal coupling \cite{HI}, and Starobinsky inflation with $R+R^2$ type action for
 gravity \cite{inf1,Starobinsky85}. It is known that Starobinsky inflation and conventional Higgs inflation
 have practically the same potential at large field values in Einstein frame, and thus lead to
 the same predictions for $\,(n_s^{},\,r)$.\, In particular, because of the exponentially flat potential
 during inflation, they predict the tensor-to-scalar ratio $\,r = \order{10^{-3}}$\,.\,
 On the other hand, our ASHI model has much flatter inflation potential than
 Starobinsky inflation and conventional Higgs inflation,
 so the energy scale of inflation is also lower.
 In consequence, our prediction of $\,r = \order{10^{-7}}$\, is smaller.
 Furthermore, as reviewed in Sec.\,1, the conventional Higgs inflation puts tight constraint
 on the Higgs mass $\,m_h^{}\,$ and top quark mass $\,m_t^{}\,$ from the requirement of
 stable inflation potential.
 This constraint pushes the required $\,(m_h^{},\,m_t^{})\,$ values beyond their
 $2\sigma$ bounds by the current collider data \cite{Higgs2014,mt-new,Top2014}.
 On the other hand, to generate the correct amplitude of CMB anisotropy,
 our ASHI model also puts nontrivial constraints on $(m_h^{},\,m_t^{})$,\,
 which fully agree with the current collider measurements
 within about 1 standard deviation, as shown in Fig.\,\ref{fig:3}.

\section{Conclusions}
\label{sec:4}
\vspace*{2mm}

 Higgs inflation is one of the most economical and predictive approaches to the cosmological inflation paradigm,
 which provides initial conditions for our universe before starting the Big Bang.

 The conventional Higgs inflation\,\cite{HI,HI2} makes use of the nonminimal coupling of the SM Higgs boson with
 Einstein general relativity.
 In this work, we constructed a new model of asymptotically safe Higgs inflation (ASHI)
 which minimally couples the SM Higgs boson to Einstein
 gravity and has the Higgs boson act as the inflaton.
 Our conjecture for the ASHI is that the Einstein gravity will exhibit
 asymptotic safety (AS)\,\cite{AS,ASReview} in the UV region, which may appear at a relatively low scale
 $\,\mutr \sim 10^{11}$\,GeV,\, around which all matter couplings become effectively zero
 (due to the Gaussian UV fixed points)\,\cite{AS_HiggsMass,AFg0,AFg}.
 In this case, the new physics is provided by quantum gravity itself.
 With these, we can achieve a fairly flat Higgs potential at the inflation scale under the impact of AS property
 on the running of Higgs self-coupling (Figs.\,\ref{fig:1}$-$\ref{fig:2}).
 We demonstrated that the model can produce correct amount of inflation, and agrees well with
 both the collider measurements on Higgs and top masses (Fig.\,\ref{fig:3}),
 and the cosmological observations (Fig.\,\ref{fig:4}).
 The inflation ends when the Higgs potential becomes steep enough, with which the energy goes down,
 and all matter couplings increase promptly from zero to the SM values.
 Thus, an efficient reheating process can successfully take place,
 driving the universe into the Big Bang era.

\vspace*{7mm}
\noindent
{\bf Acknowledgements}\\[1.5mm]
 We thank John R.\ Ellis, Yun-Long Lian, Michelangelo Mangano, Anupam Mazumdar,
 and Christoph Rahmede for useful discussions.
 This work was supported by National NSF of China (under grants 11275101, 11135003)
 and National Basic Research Program (under grant 2010CB833000).


\vspace*{3mm}
\appendix

\vspace*{3mm}
\section{Two-Loop Renormalization Group Equations for ASHI}
\label{app}
\vspace*{2mm}

{\allowdisplaybreaks

 In this appendix, we summarize the renormalization group equations
 used in our analysis of running couplings.
 For our ASHI model, we have three gauge couplings $\,(g_s^{},\,g,\,g')$\,
 for the SM gauge group $SU(3)_C^{}\otimes SU(2)_L^{}\otimes U(1)_Y^{}$,\,
 the Higgs coupling $\,\lam\,$ for the Higgs potential,
 and the Yukawa coupling $\,y_t^{}\,$ of top quark.
 (The other light fermion Yukawa couplings are negligible for our present analysis.)
 These couplings satisfy the renormalization group equations as follows,
 \bge
   \FR{\di Y}{\,\di \ln\mu\,} ~=~ \beta_{Y}^{}(g_j^{},\,\lam,\,y_t^{}) \,,
 \ede
 where $\,Y\,$ denotes one of the couplings shown above.
 For the sake of our ASHI analysis, we give all relevant $\beta$-functions up to two-loop level,
 as required for computing the Higgs potential $\,V(h)\,$.

 These beta functions coincide with that of the SM below the transition scale $\,\mutr\,$.\,
 Thus, we have the following renormalization group equations \cite{RGE},
 %
 \begin{align}
   \be_{g_s}^{}
   =&~\FR{g_s^3}{(4\pi)^2}\(-7\)+\FR{g_s^3}{(4\pi)^4}\!\(\!\FR{11}{6}g'^2+\FR{9}{2}g^2-26g_s^2-2y_t^2\)\!,
   \\[2mm]
   \be_{g}^{}
   =&~\FR{g^3}{(4\pi)^2}\(\!-\FR{19}{6}\)+\FR{g^3}{(4\pi)^4}\!
   \(\!\FR{3}{2}g'^2+\FR{35}{6}g^2+12g_s^2-\FR{3}{2}y_t^2\) \!,
   \\[2mm]
   \be_{g'}^{}
   =&~\FR{g'^3}{(4\pi)^2}\FR{41}{6}+\FR{g'^3}{(4\pi)^4}
   \(\!\FR{199}{18}g'^2+\FR{9}{2}g^2+\FR{44}{3}g_s^2-\FR{17}{6}y_t^2\)\!,
   \\[2mm]
   \be_{\lam^{}}^{}
   =&~\FR{1}{(4\pi)^2}\left[24\lam^2-6y_t^4+\FR{3}{8}
   \big(2g^4+(g^2+g'^2)^2\big)+\lam(-9g^2-3g'^2+12y_t^2)\right]
   \n\\[1.5mm]
                &~+\FR{1}{(4\pi)^4}\bigg\{\FR{1}{48}\Big(915g^6-289g^4g'^2-559g^2g'^4-379g'^6\Big)
   +30y_t^6-y_t^4\(\!\FR{8}{3}g'^2+32g_s^2+3\lam\)
   \n\\[1.5mm]
                &~+\lam\(\!-\FR{73}{8}g^4+\FR{39}{4}g^2g'^2
                +\FR{629}{24}g'^4 +108g^2\lam+36g'^2\lam-312\lam^2\)
   \n\\[1.5mm]
                &  ~+y_t^2\bigg[\!-\FR{9}{4}g^4+\FR{21}{2}g^2g'^2-\FR{19}{4}g'^4+\lam
                \(\!\FR{45}{2}g^2+\FR{85}{6}g'^2+80g_s^2-144\lam\)\!\bigg]\bigg\},
   \\[2mm]
   \be_{y_t^{}}^{}
   =&~\FR{y_t^{}}{(4\pi)^2}
      \left[-\FR{9}{4}g^2-\FR{17}{12}g'^2-8g_s^2+\FR{9}{2}y_t^2\right]
   \n\\[1.5mm]
             &~+\FR{y_t^{}}{(4\pi)^4}\bigg[\!-\FR{23}{4}g^4-\FR{3}{4}g^2g'^2
             +\FR{1187}{216}g'^4+9g^2g_s^2+\FR{19}{9}g'^2g_s^2-108g_s^4
   \n\\[1.5mm]
             &~+y_t^2\!\(\!\FR{225}{16}g^2+\FR{131}{16}g'^2+36g_s^2\)
               +6\(-2 y_t^4-2 y_t^2\lam+ \lam^2\)\!\bigg].
 \end{align}
 %



\vspace{9mm}
%

\end{document}